\documentclass[apj,iop]{emulateapj}
\usepackage{natbib}
\usepackage{tablefootnote}
\newcommand{\dnu}{$\Delta\nu$}
\newcommand{\numax}{$\nu_{\text{max}}$}
\newcommand{\Teff}{$T_{\mathrm{eff}}$}
\newcommand{\uhz}{$\mu \textrm{Hz}$}
\newcommand{\chisq}{$\chi^2$}
\newcommand{\dia}{\textsc{diamonds}}

\newcommand{\kepler}{\textsl{Kepler}}
\newcommand{\Msol}{$M_{\odot}$}
\newcommand{\Rsol}{$R_{\odot}$}
\newcommand{\cref}{\textbf{(reference?)}}

\shorttitle{Helium Abundance of NGC 6791}
\shortauthors{McKeever et al.}

\begin{document}
\title{The Helium Abundance of NGC 6791 from Modeling of Stellar Oscillations}

\author{Jean M. McKeever and Sarbani Basu}
\affil{Astronomy Department, Yale University,
    New Haven, CT 06511}

\author{Enrico Corsaro}
\affil{INAF -- Osservatorio Astrofisico di Catania, via S. Sofia 78, 95123 Catania, Italy}

\email{jean.mckeever@yale.edu}

%\altaffiltext{1}{Alternate affiliation.}

\begin{abstract}

The helium abundance of stars is a strong driver of evolutionary timescales, however it is difficult to measure in cool stars. We conduct an asteroseismic analysis of NGC 6791, an old, metal rich open cluster that previous studies have indicated also has a high helium abundance. The cluster was observed by \kepler\ and has unprecedented lightcurves for many of the red giant branch stars in the cluster. Previous asteroseismic studies with \kepler\ data have constrained the age through grid based modeling of the global asteroseismic parameters (\dnu\ and \numax). However, with the precision of \kepler\ data, it is possible to do detailed asteroseismology of individual mode frequencies to better constrain the stellar parameters, something that has not been done for these cluster stars as yet. In this work, we use the observed mode frequencies in 27 hydrogen shell burning red giants to better constrain initial helium abundance ($Y_0$) and age of the cluster.  The distributions of helium abundance and age for each individual red giant are combined to create a final probability distribution for age and helium abundance of the entire cluster. We find a helium abundance of $Y_0=0.297\pm0.003$ and a corresponding age of $8.2\pm0.3$ Gyr.
\end{abstract}

\keywords{open clusters: general;
open clusters: individual (\objectname{NGC 6791}); stars: oscillations}

\section{Introduction}
As the second most abundant element, helium provides valuable insight into many astrophysical situations. The helium abundance of stars has a direct link to age, such that stars evolve more quickly, and at a higher temperature and luminosity, with a larger helium abundance. This leads to a red giant branch that is hotter than one with lower helium content \citep{Salaris2006}. The study of helium can also allow us some insight into chemical evolution in the galaxy. Helium abundance is particularly difficult to measure in stars cooler than 12000 K, where the lines are very weak. Helioseismic measurements of helium abundance for the Sun \citep{Dappen1991,Vorontsov1991,Basu1995,Basu2004}, 16 Cyg \citep{Verma2014}, and HD176465 \citep{Gai2018} are the only individual stellar measurements of helium abundance in cooler stars, although a few others have been analyzed (Verma et al., submitted). %Many stellar models and isochrones are created with fixed value of $\Delta Y/\Delta Z$, however it is not well constrained.

Stellar clusters are a unique setting in which all members share the same age and composition. This enables us to study many individual stars to determine cluster properties with better precision.  NGC 6791 is an old ($\sim8$ Gyr) and metal rich ([Fe/H]$\sim+0.4$) open cluster. The combination of old age and high metallicity makes it a very interesting place to study the helium abundance. There have been many studies of the cluster and extensive work has been done to study the age of the cluster through various techniques including isochrone fitting \citep{Harris1981,Anthony-Twarog1985,Stetson2003}, binaries \citep[][hereafter B2012]{Brogaard2011,Brogaard2012}, white dwarfs \citep{Bedin2005,Bedin2008,Garcia-Berro2010}, and asteroseismology \citep{Basu2011}, however, not much has been done regarding helium specifically.  \citet{Brogaard2011}, as part of their analysis,  had noted that the helium abundance of the cluster from their isochrones was likely super-solar, with a value of $Y_0\sim0.30$.

Asteroseismology is a unique tool to study the interiors of stars and derive fundamental parameters, such as mass and radius, with very high precision. In conjunction with a set of reasonable stellar models, we are able to determine the properties of stars, including ages, to very high precision using the information contained in the oscillations. Thus, asteroseismic modeling of stars in clusters provides an ideal method to examine the helium abundance of the cluster in detail, especially for cool stars where traditional methods such as spectroscopy are limited.

In NGC 6791 specifically, \citet{Hekker2011NGC6791} examined the global properties of giants in the cluster, such as the mass and radius distributions, through an asteroseismic analysis of the red giants, using only the global asteroseismic parameters. By using some of the extra information contained in individual frequencies, \citet{Corsaro2012} determined the period spacing of observed modes as a way to distinguish between red clump and red giant branch stars.  However, there as yet are no studies that approach the cluster by modeling the oscillations of individual stars directly.

%--about 10\% for main sequence stars \citep{SilvaAguirre2016}--

 %The resultant power spectrum of these oscillations is very regular in the spacing of the mode frequencies and can be easily described by the two global oscillation parameters: the large frequency spacing, \dnu, which is the spacing between mode frequencies of the same degree, and the frequency of maximum power, \numax.

Modeling of the individual oscillation modes in stars across the HR diagram is not a new idea, however until recently the observational data did not exist. With new space-based satellites such as CoRoT \citep{Baglin2006} and \kepler\ \citep{Borucki2010}, solar-like oscillations were exposed in thousands of red giants \citep{Hekker2009,Kallinger2010corot,kallinger2010kep,Bedding2010,Huber2010,Hekker2011}. The potential for the discovery of many more oscillating stars exists with current and forthcoming missions such as TESS \citep{Ricker2015} and PLATO \citep{Rauer2014}. Solar-like oscillations arise in stars with outer convective layers. The oscillations are stochastically driven by the turbulent convection in the outer layers. There are several examples of asteroseismic modeling applied to main sequence stars \citep{Miglio2005,Lebreton2014,Metcalfe2014,Appourchaux2015,Roxburgh2016,SilvaAguirre2017,Creevey2017,White2017,Bazot2018},  and recently to several red giants \citep{Miglio2010,PerezHernandez2016,Triana2017,Ball2018}.

%Age and helium abundance have some degeneracy between them with the more helium rich stars being typically younger, and as such, helium is important to take into account.  %As you move up the red giant branch, the density of mixed modes increases and identification of dipole modes, in particular, is much more complicated. In this work we will only make use of the radial modes and quadrupole modes, where we typically do not observe mixed modes as the inertia of the mixed modes is too large to have an impact on the quadrupole frequency. For this reason we select the mode of lowest inertia per radial order for $l=2$

In this work we examine the helium abundance in NGC 6791 through two different approaches. First, we model the eclipsing binaries of B2012, taking into account a wide range of initial helium abundances in a thorough search through a fine grid of stellar evolution models. And secondly, with an asteroseismic study of the red giants in the cluster. The subgiants and main sequence stars of NGC 6791 are too faint for asteroseismic detections with \kepler. We fitted the oscillation frequencies of 27 red giants and match them to frequencies computed from stellar evolution models. Finally, we use the fact that as cluster members, all stars should have the same age and initial composition to do a joint analysis of both helium abundance and age. 

The  rest of our paper is organized as follows: In Section~\ref{sec:data} we layout our target selection as well as describe the global asteroseismic parameter determination and the individual frequency fitting. In Section~\ref{sec:models} we explain the range of parameters and choice of physics included in our models. Section!\ref{sec:results} compares our models along the main sequence to the results obtained by B2012 %\citet{Brogaard2012a}, 
and we discuss our results obtained from the red giants.  We conclude with a short summary in Section \ref{sec:discussion}.

\section{Data \& Target Selection}
\label{sec:data}
\subsection{Eclipsing Binaries}
We used the two eclipsing binary star systems in the cluster previously studied by B2012 %\citet{Brogaard2012a}
to test our models and confirm their results. They span a large range of masses along the main sequence up to the turnoff; their properties are summarized from B2012 %\citet{Brogaard2012a}
in Table~\ref{tab:binary}. The secondary star in V20 does not have an independently measured metallicity, so we assumed [Fe/H]$=0.26$, the same as the primary. There is no asteroseismic data for these stars, however, as we shall show, the dynamical and spectroscopic parameters are sufficient to constrain the helium abundance. 

\begin{deluxetable*}{ccccc}
\label{tab:modes1}
%\tablecolumns{4}
%\tablewidth{0pt}
%\tabletypesize{\small}
\tablecaption{Binary star parameters\label{tab:binary}}
\centering
\tablehead{
  \colhead{}& \colhead{V18}& \colhead{V18} & \colhead{V20}& \colhead{V20}\\
  \colhead{}& \colhead{Primary}& \colhead{Secondary} & \colhead{Primary}& \colhead{Secondary}}\\
\startdata
Mass [\Msol] &  0.9955$\pm$0.0033 & 0.9293$\pm$0.0032 & 1.0868$\pm$0.0039 & 0.8276$\pm$0.0022 \\
Radius [\Rsol]& 1.1011$\pm$0.0068 & 0.9708$\pm$0.0089 & 1.397$\pm$0.013 & 0.7813$\pm$0.0053 \\
Teff [K] & 5600$\pm$95 & 5430$\pm$125 & 5645$\pm$95 & 4860$\pm$125 \\
$[\textrm{Fe/H}]$& 0.31$\pm$0.06 & 0.22$\pm$0.10 & 0.26$\pm$0.06 & \nodata
\enddata
\tablecomments{For the purpose of fitting, the secondary of V20 was assumed to have the same metallicity as the primary component of V20}
\end{deluxetable*}

\subsection{Red giant data}
\subsubsection{Asteroseismic data}
\begin{figure}
%\epsscale{.80}
\plotone{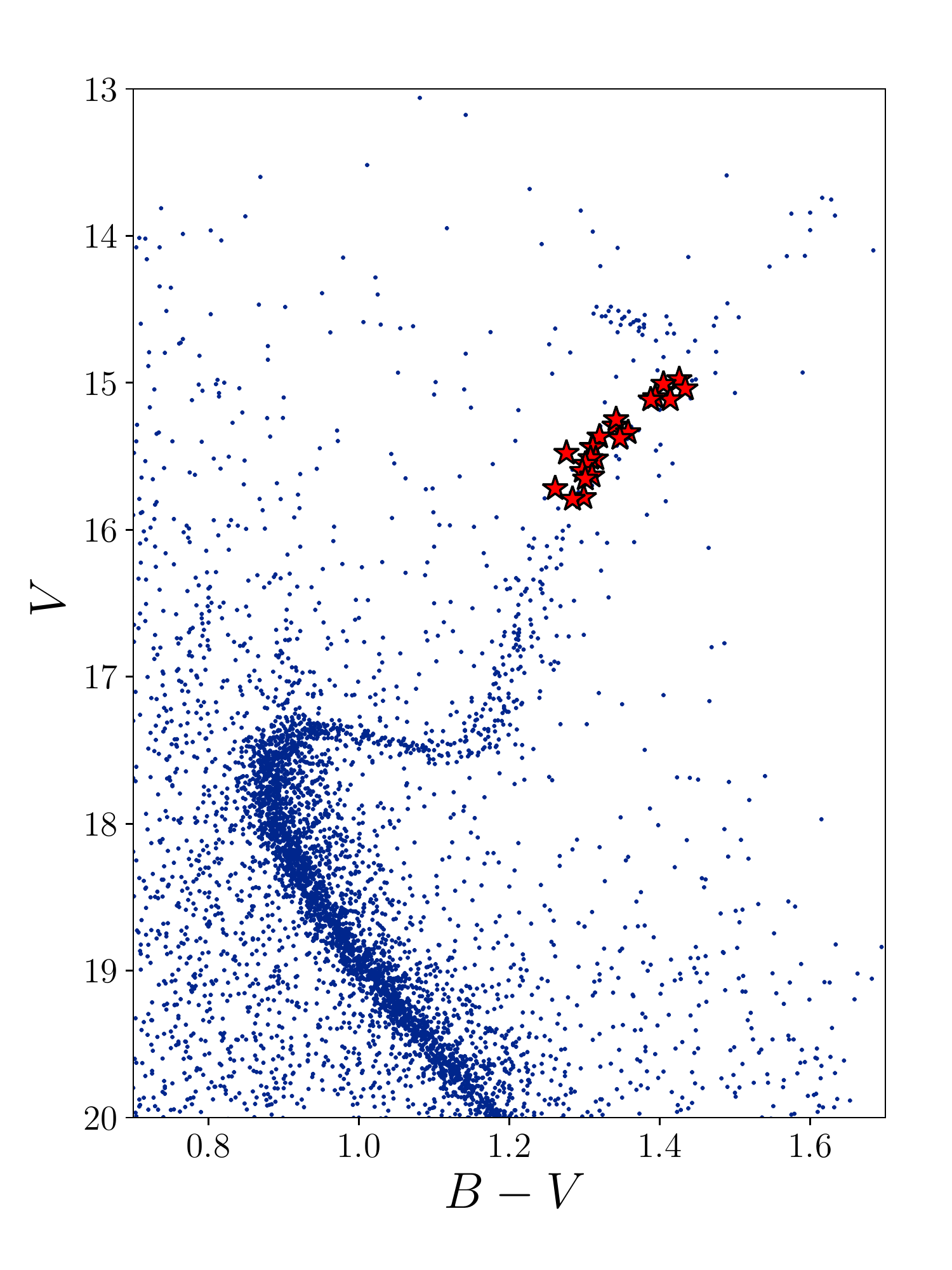}
\caption{Color-Magnitude Diagram of NGC 6791 with photometry of \citet{Stetson2003}. Red stars denote targets studied in this work. See Table~\ref{tab:star_props} for a list of their general properties.\label{fig:CMD}}
\clearpage
\end{figure}

NGC 6791 was observed for four years with the \kepler\ satellite, which has provided remarkable light curves for many of the stars in the field. We retrieved long cadence (30 min) light curves for Q1-Q17 from the Kepler Asteroseismic Science Operations Center (KASOC)\footnote{http://kasoc.phys.au.dk/} which produces light curves and power spectra that have been processed for asteroseismology specifically \citep{Handberg2014}.  From the red giants previously identified in the cluster \citep{Stello2011}, we chose stars with a good signal-to-noise ratio and clearly visible modes in the power spectrum.  The stars chosen for this study are highlighted in Fig.~\ref{fig:CMD}, which shows a color-magnitude diagram of the cluster with the photometry of \citet{Stetson2003}. Their global asteroseismic parameters and available spectroscopic data are listed in Table~\ref{tab:star_props}. We considered only  stars on the red giant branch. %To model all the stars efficiently, only first ascent red giants are considered, as both the evolution models and the oscillation calculations take much longer on the horizontal branch.%, however there are oscillating for core helium burning stars. 

\begin{deluxetable*}{cccccc}
\label{tab:modes1}
%\tablecolumns{4}
%\tablewidth{0pt}
%\tabletypesize{\small}
\tablecaption{Global asteroseismic and spectroscopic properties of target red giants.  \label{tab:star_props}}
\centering
\tablehead{
  \colhead{KIC}& \colhead{\numax}& \colhead{\dnu}& \colhead{\Teff}  & \colhead{[Fe/H]$^a$}& \colhead{RV$^a$}\\
    \colhead{}& \colhead{\uhz}& \colhead{\uhz}& \colhead{K}  & \colhead{dex} & \colhead{km/s}}
\startdata
%2437507  &   20.43$\pm$0.09  &  2.61$\pm$0.03  & 4246$\pm$100  &    &  \\ 
%2569360  &   21.70$\pm$0.08  &  2.72$\pm$0.03  & 4254$\pm$100  &    &   \\
2436814  &   25.56$\pm$0.04  &  3.11$\pm$0.03  & 4289$\pm$100  &    &   \\
%2436332  &   28.69$\pm$0.09  &  3.39$\pm$0.29  & 4304$\pm$100  &    &   \\
2436824  &  34.26$\pm$0.13  &  3.85$\pm$0.04  & 4324$\pm$100 &    &   \\
2436900  &  35.67$\pm$0.18  &  4.02$\pm$0.04  & 4403$\pm$78  & 0.40$\pm$0.02   & -48.37  \\
2436458  &  35.87$\pm$0.24  &  4.14$\pm$0.03  & 4340$\pm$100  &    &   \\
2435987  &  36.26$\pm$0.20  &  4.19$\pm$0.04  & 4434$\pm$100 & 0.38$\pm$0.02   & -43.75  \\
2436097  &  40.53$\pm$0.22  &  4.54$\pm$0.04  & 4365$\pm$100 &    &   \\
2437240  &  45.58$\pm$0.22  &  4.87$\pm$0.05  & 4440$\pm$100 &    &   \\
2437402  &  46.09$\pm$0.26  &  4.82$\pm$0.05  & 4414$\pm$100 &    &   \\
2570518  &  46.48$\pm$0.16  &  4.94$\pm$0.05  & 4496$\pm$71  & 0.38$\pm$0.02   & -48.88  \\
2569618  &  56.01$\pm$0.16  &  5.68$\pm$0.07  & 4479$\pm$81  & 0.40$\pm$0.02   & -46.12  \\
2436540  &  57.31$\pm$0.28  &  5.82$\pm$0.07  & 4492$\pm$83  & 0.41$\pm$0.02   & -48.60  \\
2436209  &  57.62$\pm$0.24  &  5.76$\pm$0.08  & 4498$\pm$83  & 0.43$\pm$0.02   & -48.31  \\
2438333  &  61.07$\pm$0.14  &  6.11$\pm$0.08  & 4522$\pm$80  & 0.43$\pm$0.02   & -48.10  \\
%2570384  &  61.49$\pm$0.89  &  6.39$\pm$0.15  & 4553$\pm$79  & 0.39$\pm$0.02   & -47.97  \\
2438038  &  62.56$\pm$0.21  &  6.13$\pm$0.07  & 4450$\pm$100 &    &   \\
2437488  &  65.30$\pm$0.19  &  6.31$\pm$0.09  & 4452$\pm$100  &    &   \\
2570094  &  68.39$\pm$0.25  &  6.45$\pm$0.08  & 4485$\pm$100 &    &   \\
2438140  &  71.37$\pm$0.24  &  6.72$\pm$0.10  & 4543$\pm$100  &    &   \\
2437653  &  74.20$\pm$0.26  &  6.96$\pm$0.10  & 4588$\pm$83  & 0.42$\pm$0.02   & -46.89  \\
2570172  &  74.33$\pm$0.28  &  7.00$\pm$0.09  & 4536$\pm$85  & 0.44$\pm$0.02   & -47.24  \\
2436688  &  76.06$\pm$0.39  &  7.22$\pm$0.10  & 4537$\pm$86  & 0.43$\pm$0.02   & -47.89  \\
2437972  &  84.66$\pm$0.29  &  7.84$\pm$0.13  & 4543$\pm$100 &    &   \\
2437781  &  85.15$\pm$0.29  &  7.74$\pm$0.16  & 4456$\pm$100  &    &   \\
2437976  &  89.83$\pm$0.35  &  8.16$\pm$0.15  & 4525$\pm$100  &    &   \\
2437957  &  92.71$\pm$0.38  &  8.36$\pm$0.20  & 4602$\pm$100  &    &   \\
2437325  &  93.50$\pm$0.35  &  8.45$\pm$0.21  & 4557$\pm$100  &    &   \\
2570244  &  106.31$\pm$0.33  &  9.17$\pm$0.27  & 4559$\pm$100  &    &   \\
2437933  &  107.72$\pm$0.33  &  9.39$\pm$0.32  & 4610$\pm$100  &    &

\enddata 
\footnote[0]{$^{a}$Only stars with APOGEE data have metallicity and radial velocity listed} 
\clearpage
\end{deluxetable*}

As a preliminary estimate on the global asteroseismic parameters, the large separation (\dnu) and the frequency of maximum power (\numax), were calculated from a 2D auto-correlation \citep{Mosser2009}. The value of \numax\ was further refined during our background fitting as part of the process of measuring mode frequencies. We used the \dia\footnote{https://github.com/EnricoCorsaro/DIAMONDS} code \citep{Corsaro2014,Corsaro2015} to fit the background of the power spectrum. \dia\ uses a nested sampling Monte Carlo method to efficiently arrive at a best model given a Bayesian prior. The power in the regions away from the acoustic modes can be described as a combination of three semi-Lorentzians \citep{Kallinger2014}, 

\begin{equation}
\label{eq:bg}
B(\nu)=\sum\limits_{i=1}^{3}\frac{\xi\sigma_i^2/\nu_{0,i}}{1+(\nu/\nu_{0,i})^4}+W,
\end{equation}
with the parameters $\nu_{0,i}$ representing the characteristic time scale of long-term photometric variability and granulation on two size scales,  $\sigma_i$ is the amplitude of each of those components, $\xi$ is a normalization term of the semi-Lorentzian equal to $2\sqrt{2}/\pi$,  and a white noise parameter, $W$.

\begin{figure*}
\epsscale{1.55}
\plottwo{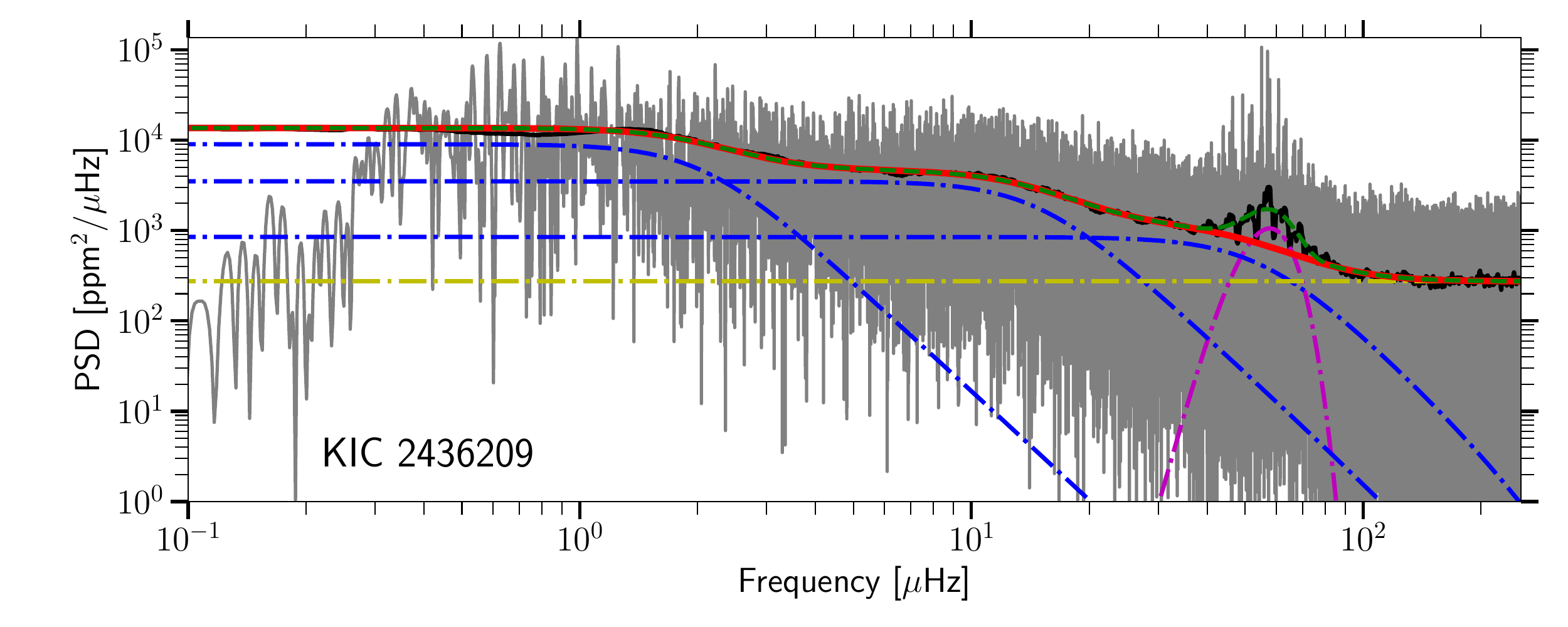}{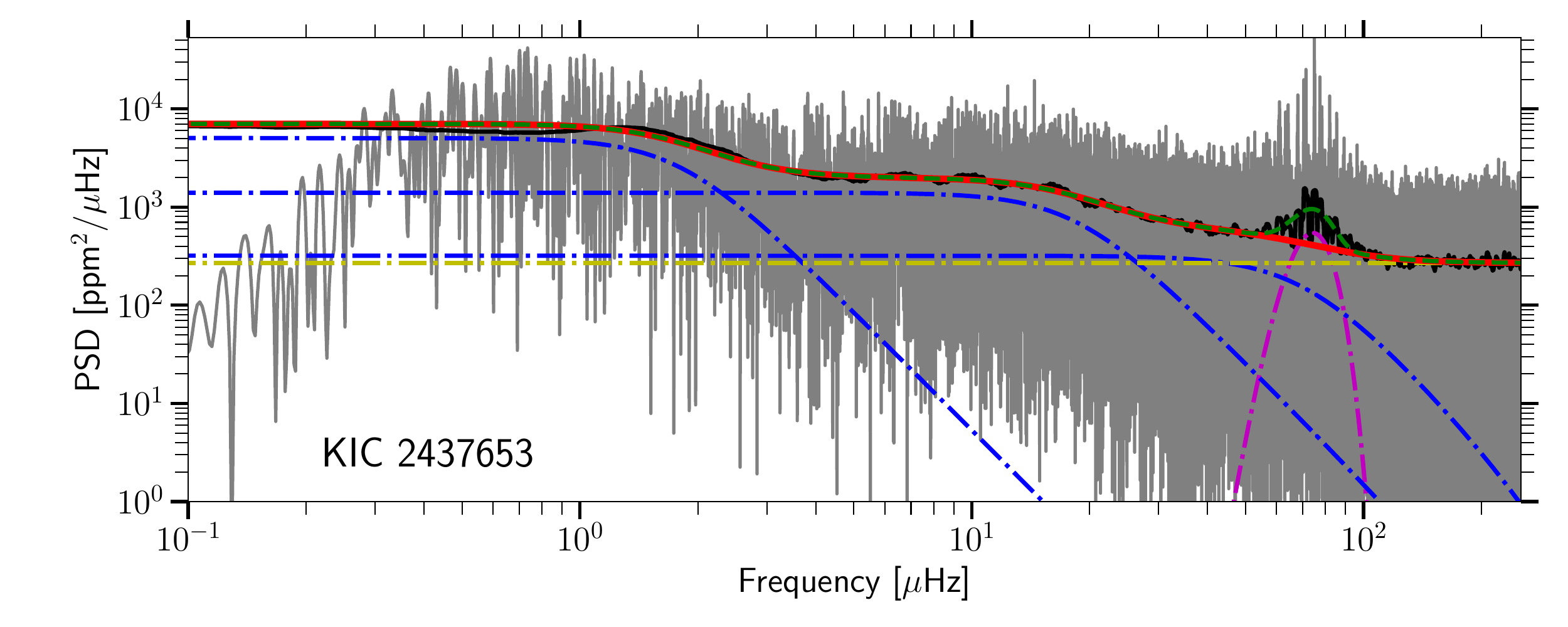}
\plottwo{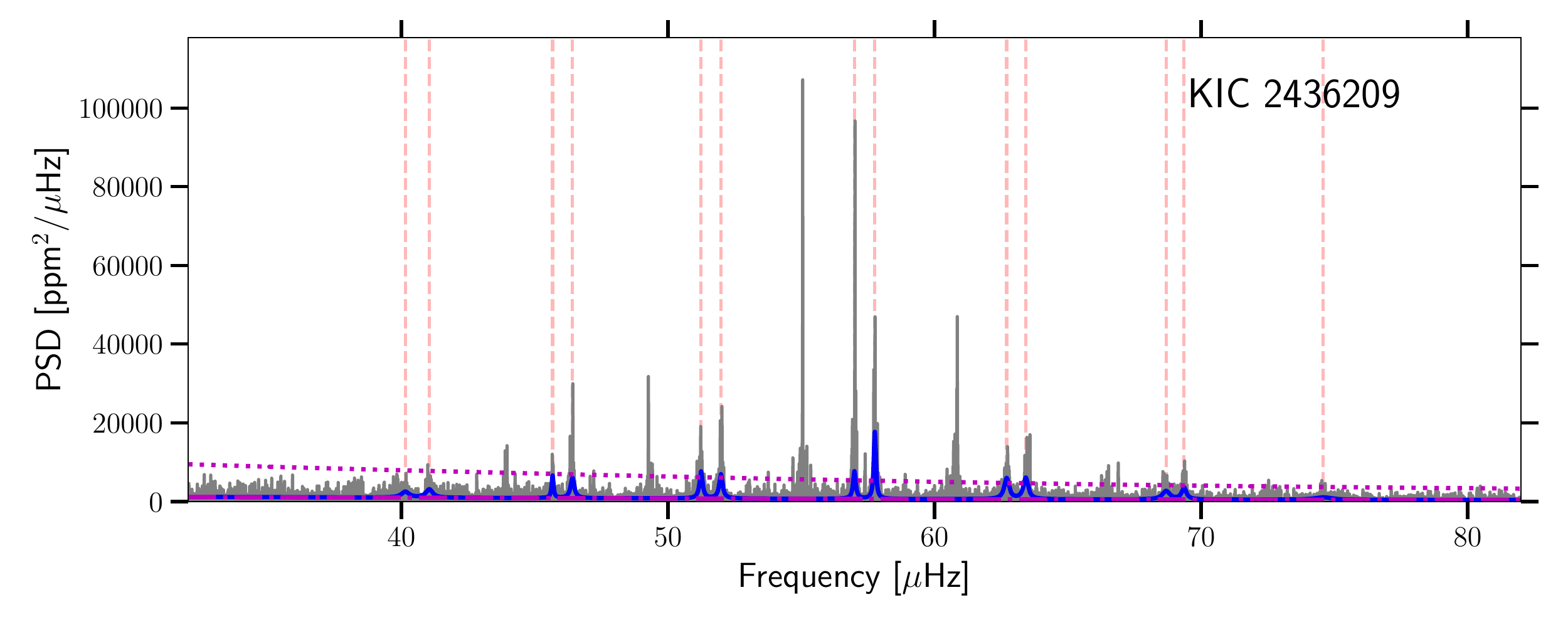}{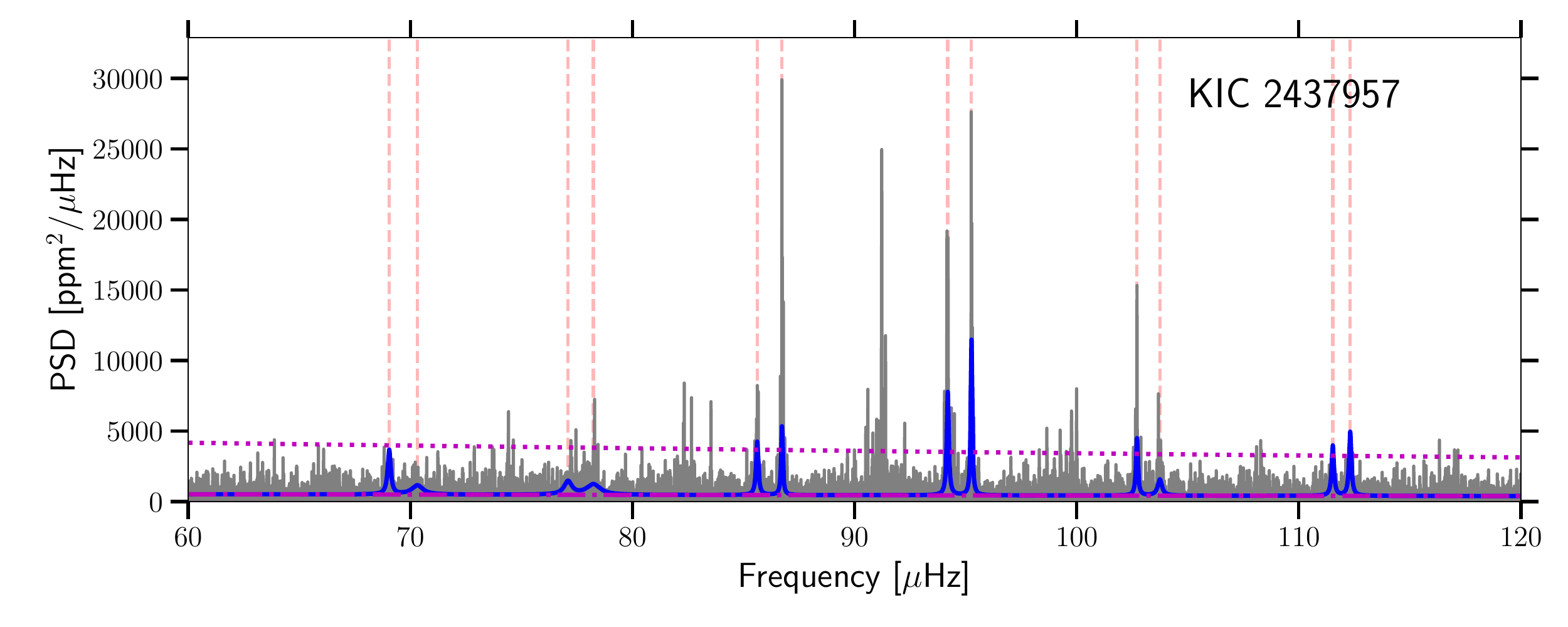}
\caption{\textsl{Top two}: Examples of the fitted background determined by \dia. The observed power spectrum is shown in gray and a smoothed version in black. The total fitted background is shown with (green dashed line) and without (red line) the Gaussian component included. The individual components are drawn with blue (semi-Lorentzians), yellow (white noise level), and magenta (Gaussian) lines.  \textsl{Bottom two}: Results of mode fitting for two targets. Again, the power spectrum is in gray and the fitted peaks are shown in blue. Vertical red dashed lines are placed at the locations of the central frequency of each mode. The back ground level (dashed) and eight times the background level (dotted) are both shown in magenta.  \label{fig:dia_BG}}
\clearpage
\end{figure*}

The envelope of the mode excess is approximated as a Gaussian, 
\begin{equation}
\label{eq:gauss}
G(\nu)=H_0\exp{\left[-\frac{(\nu-\nu_{\rm max})^2}{2\sigma_g^2}\right]},
\end{equation}
where $H_0$ is the amplitude, \numax\ is the frequency at maximum amplitude, and $\sigma_g$ is the the standard deviation of the Gaussian, related to the width of the observable mode envelope.  Including the Gaussian function when fitting the background ensures that the background level is accurately measured in the region of the power excess where the modes would otherwise raise the background level. The top two panels of Fig.~\ref{fig:dia_BG} show the fitted background for two of our targets.

Individual modes frequencies were also fitted using \dia. The methodology of using \dia\ to fit red giants has been described by \citet{Corsaro2015,Corsaro2018}. If a mode has a lifetime sufficiently shorter than the duration of the observations, then the mode is considered resolved and has a Lorentzian profile in the power spectrum, given as
\begin{equation}
P(\nu)=\frac{A^2/(\pi\Gamma)}{1+4/\Gamma^2(\nu-\nu_0)^2},
\end{equation}
where $A$ is the amplitude, $\nu_0$ is the frequency of the mode, and $\Gamma$ is the linewidth, which is inversely related to the mode lifetime. 

The mode frequencies were also extracted by \citet{Corsaro2017Spin} for many of the targets, however, for consistency all targets were reanalyzed in this work. Very good agreement was found between both sets of frequencies. %We only consider resolve radial ($l=0$) and quadrupole ($l=2$) modes in this work. %Unresolved modes do not have measurable linewidths and are typically fit as a sinc-squared function.% Pure p-modes have shorter lifetimes compared to the g-modes propagating in the core. This leads to mixed modes having a more unresolved structure. 
For this study we are using only the radial ($l=0$) and quadrupole ($l=2$) modes which can be treated as resolved modes. There are very few published period spacings for our target stars \citep{Mosser2018}; therefore, we chose to exclude the period spacing as an additional constraint to maintain a homogeneous data set. The bottom two panels of Fig.~\ref{fig:dia_BG} show the results of the mode fitting process for two of our targets. %The dipole ($l=1$) modes are ignored for the moment due to the complicated structure of the mixed modes making individual mode identifications much more challenging. %Table \ref{tab:freqs} (available in its entirety online) lists our the fitted frequencies.

\subsubsection{Spectroscopic data}
Elemental diffusion in stars leads to a spread in surface metallicity along evolutionary tracks, with red giants having typically higher surface metallicity than main sequence stars in a single simple stellar population (see Fig.~7 of B2012). This can be seen in previous spectroscopic studies of the cluster. For main sequence binary stars \citet{Brogaard2011} find [Fe/H] $=0.22$--$0.34$. \citet{Boesgaard2009} measured [Fe/H] $=0.30$ in turn-off stars of the cluster. For M-giants and red giants, respectively, \citet{Origlia2006} measure [Fe/H] $=0.35$ and \citet{Carraro2006} find [Fe/H] $=0.39$. There is a clear pattern showing higher metallicity at more evolved states.  

In addition, many of the stars in the cluster were observed by APOGEE \citep{Majewski2017}, a spectroscopic survey largely targeting red giants throughout the galaxy. \citet{Corsaro2017} used APOGEE DR13 \citep{Albareti2017} data in their analysis of granulation properties and found an average [Fe/H]=0.32. We use the spectroscopic metallicities and temperatures from DR14 \citep{Abolfathi2018}, which includes some updates to the analysis pipeline \cite{Holtzman2018}, as constraints in our models. Both the radial velocities and metallicities reported by APOGEE from their ASPCAP pipeline are very homogeneous for stars in the cluster, with a mean [Fe/H]=0.41. As such, we adopt a metallicity for stars without data that is the mean metallicity of stars with APOGEE data. Temperatures, where available, were also taken from APOGEE. For stars without APOGEE data,  temperatures were taken from \citet{Basu2011}, where they derived $(V$ -- $K)$ color-based temperatures from existing photometry.

When studying individual mode frequencies it is important to correct for the frequency shifts caused by the relative motion of the Earth around the Sun \citep{Davies2014}. While the effect is small, with high-precision data such as \kepler\ lightcurves, it can be on the order of your typical errorbars. For these corrections, radial velocities from APOGEE were used where available and a mean APOGEE radial velocity adopted for those without.

\section{Stellar Models and Pulsation Calculations}
\label{sec:models}
We used the stellar evolution code Modules for Experiments in Stellar Astrophysics \citep[MESA\footnote{http://mesa.sourceforge.net/}, v10108][]{Paxton2011,Paxton2013,Paxton2015,Paxton2018} to create our models. We chose the abundance mixtures to be those of \cite{Grevesse1998}.  High-temperature opacities come from OPAL \citep{Iglesias1993,Iglesias1996} while low temperatures are covered by \citet{Ferguson2005}. The EOS also comes from OPAL \citep{Rogers2002}, and is the default option given in MESA (see Paxton et al. 2011 for further details). The nuclear reactions are from NACRE group \citep{Angulo1999} and supplemented by \citet{Caughlan1988} with some updates to $^{12}$C$(\alpha,\gamma)^{16}$O \citep{Kunz2002} and $^{14}$N$(p,\gamma)^{15}$O \citep{Imbriani2005}. %, and triple-$\alpha$ \citep{Fynbo2005}. 
The models consider convection as described by the mixing-length theory \citep{Cox1968}  and  the gravitational settling of elements \citep{Thoul1994}. A small amount of convective overshoot as described by \cite{Herwig2000} was allowed during both the main sequence and red giant phases.
%It should be noted that for our range of parameters, convective core overshoot is not expected to play a significant role as the masses are generally too low to have a convective core along the main sequence. 
We used the Eddington gray atmosphere approximation for the outer boundary condition. 

We allowed four parameters to vary in the models we created: mass, initial [Fe/H], initial helium abundance ($Y_0$),  and mixing-length parameter $\alpha$. The exponential convective overshoot parameter was fixed at $f_0=0.016$, based on the values used in the MIST isochrones \citep{Choi2016}. %however most of the models are below where we expect convective mixing on the main sequence to play a significant role in evolution. 
Initial metallicities are relative to our chosen solar abundance, with $Z/X_{\odot} = 0.023$ \citep{Grevesse1998}.

Two grids were computed for the cluster; the first was over a mass range of the binary stars identified in B2012 (see Table~\ref{tab:binary}) and were only evolved to the onset of hydrogen shell burning. Mass points were chosen to be near the binary mass estimates. The models spanned a range in initial helium abundance of $0.26$--$0.34$. Because of the inclusion of diffusion in our models, which changes the surface metallicities, we also kept the range of initial metallicities fairly broad. The range in mixing length $\alpha$ covers 1.6 to 2.2 and encompasses our solar calibrated value of $\alpha_\odot=1.79$.

A second grid of models was created over the mass range of the red giants inferred from the asteroseismic scaling relations ($1.1$--$1.25$\Msol), and with a narrower scope of parameters than the main sequence models informed by results of the binary star comparisons. %The lower limit of 1.1\Msol is in agreement with B2012 who quote a mass of 1.08\Msol for their highest mass star which is approaching, but not yet at turn off. 
All models were evolved up the red giant branch far enough to encompass the $\log g$ range of interest. Table~\ref{tab:grids} gives an overview of the range of parameters used. 

\begin{deluxetable}{lcc}
\label{tab:modes1}
%\tablecolumns{4}
%\tablewidth{0pt}
%\tabletypesize{\small}
\tablecaption{Stellar model parameters\label{tab:grids}}
\centering
%\tablehead{
%  \colhead{Parameter}& \colhead{Range}& \colhead{Step size}& }
\startdata
Parameter & Range& Step size \\
\cutinhead{Main sequence models}
$M$  & 0.8 $\rightarrow$ 1.12  & 0.003 $\rightarrow$ 0.02  \\
$Y$  &  0.26 $\rightarrow$ 0.34&  0.005, 0.01 \\
Fe/H  & 0.15 $\rightarrow$ 0.55 &  0.05\\
$\alpha$  &  1.6 $\rightarrow$ 2.2 &  0.1\\
\cutinhead{Red giant models}
$M$  & 1.1 $\rightarrow$ 1.25  &  0.01 \\
$Y$  &  0.28 $\rightarrow$ 0.32& 0.005, 0.01\\
Fe/H  & 0.25 $\rightarrow$ 0.45 &  0.05\\
$\alpha$  & 1.7 $\rightarrow$ 2.1 &  0.1
\enddata
\clearpage
\end{deluxetable}

\begin{figure}
\epsscale{1.05}
\plotone{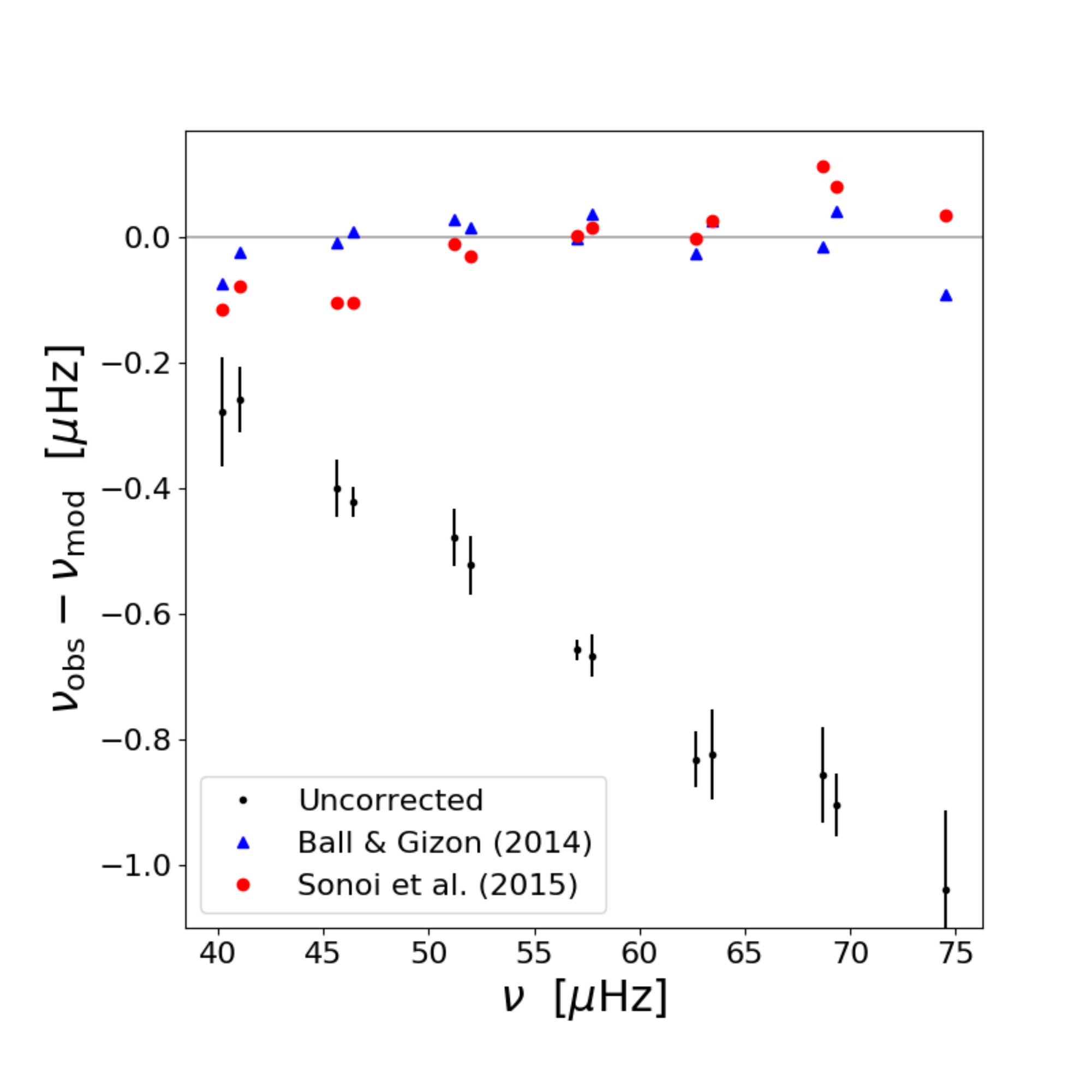}
\caption{The difference between the observed frequencies and model frequencies (black), \citet{Ball2014} corrected frequencies (blue triangles), and \citet{Sonoi2015} corrected frequencies (red circles). The errorbars are shown for the uncorrected values only. Note the frequency dependent trend left in the \citet{Sonoi2015} corrections.} \label{fig:surf_corr}
\clearpage
\end{figure}

Adiabatic mode frequencies were computed up to the Nyquist frequency for long cadence \kepler\ data (283 $\mu$Hz) for red giant models using GYRE\footnote{https://bitbucket.org/rhdtownsend/gyre/wiki/Home} \citep{Townsend2013}. Frequencies were only calculated for models where  $2.15 < \log g < 3.05$, and only for $l=0$ and $l=2$ modes. %The stellar mesh was re-sampled at a higher density to better resolve the g-mode frequencies when computing the quadrupole modes.  
Model frequencies were corrected with the two-term surface corrections of \citet{Ball2014}. The surface corrections of \citet{Sonoi2015} were also considered, however the resulting fits to the frequency difference showed a remaining frequency-dependent trend in the residuals. This is exemplified in Fig.~\ref{fig:surf_corr}, where we show the frequency differences between the observations and a near-optimal model for one of our targets corrected according to both \citet{Ball2014} and \citet{Sonoi2015}. % to account for near surface inaccuracies that we can not yet model well. %before contributing to the \chisq. 

The goodness of fit for each model in the grid is measured as a \chisq\ value that is a combination of individual \chisq\ values from the available observational constraints. For the binary stars, 
\begin{equation}
    \chi^2 = \chi^2_M + \chi^2_R + \chi^2_{T_{\mathrm{eff}}} + \chi^2_{\mathrm{[Fe/H]}},
\end{equation}
where only the spectroscopic temperature and metallicity, and the mass and radius from the binary analysis of B2012 were considered. For the red giants in our sample,
\begin{equation}
    \chi^2 = \frac{1}{3} \left ( \chi^2_{\mathrm{ast}} + \chi^2_{T_{\mathrm{eff}}} + \chi^2_{\mathrm{[Fe/H]}} \right ),
\end{equation}
where the spectroscopic constraints are those from Table~\ref{tab:star_props}, and
\begin{equation}
    \chi^2_{\mathrm{ast}} = \frac{1}{N} \sum_i \frac{(\nu_{\mathrm{obs},i} - \nu_{\mathrm{model},i})^2} {\sigma_i^2},
\end{equation}
where $N$ is the total number of observed modes with an observed frequency $\nu_{\mathrm{obs}}$, and a corresponding model frequency, $\nu_{\mathrm{model}}$.
%For main sequence models, the observables include the mass and radius from the eclipsing binary analysis, as well as the spectroscopic constraints of  metallicity and \Teff. Models computed along the red giant branch include the metallicity, \Teff, and the individual mode frequencies. 
For the red giants a cutoff of model \numax\ within 3\dnu\ of the observed \numax\ was implemented.  

A probability distribution for each free parameter of the model plus age was generated and weighted appropriately by the likelihood, which is proportional to  $\exp(-\chi^2/2)$. However, we can take advantage of the fact that these stars are all members of the same cluster and reduce the errors on our results by generating a joint probability distribution that is the product of each individual star. Furthermore, we know that the helium abundance and age are linked, so we create a 2-d probability distribution using a weighted kernel density estimator, again weighted by the likelihood.

\section{Results and Discussion}
\label{sec:results}
\subsection{Binary stars along the main sequence}
\begin{figure}
\epsscale{1.25}
\plotone{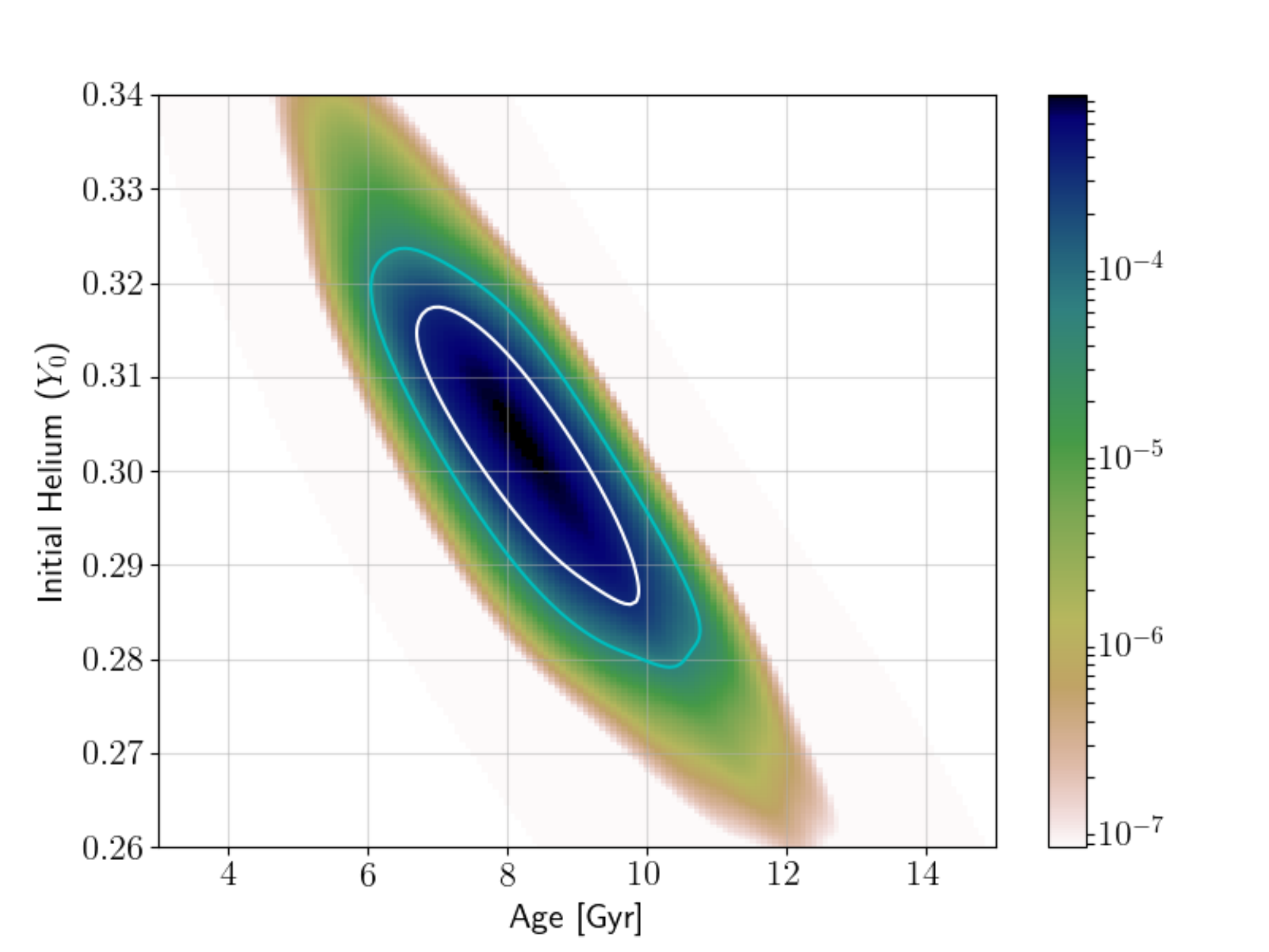}
\caption{Joint probability distribution of age and initial helium abundance for the two eclipsing binary systems of B2012. %\cite{Brogaard2012}. 
The 1-$\sigma$ and 2-$\sigma$ error contours are drawn in white and cyan, respectively. \label{fig:binary_2d}}
\clearpage
\end{figure}

We first use the binary stars of B2012 %\citet{Brogaard2012a} 
to determine the helium abundance and age of the cluster. In calculating \chisq\ for our models we assumed a [Fe/H] for the secondary of V20 that was equal to that of the primary component. We find an age of $8.5\pm1.1$ Gyr  and a $Y_0 = 0.299\pm0.011$ from the joint probability distributions of all four stars considered. Fig.~\ref{fig:binary_2d} shows the 2-d probability density function that was computed using weighted kernel density estimates. White and cyan contours indicate the 1-$\sigma$ and 2-$\sigma$ error ellipses, respectively. One can clearly see that there is a trend between helium abundance and age, where a higher helium abundance typically indicates a younger age. 

For reference, B2012 %\citet{Brogaard2012a}
determined the age to be $8.3\pm0.3$ Gyr and $Y_0 = 0.30\pm0.01$. Our results are entirely in agreement with B2012 %\citet{Brogaard2012a}
and present good supporting evidence for a super-solar helium abundance in the cluster. These results also provide a check on the consistency of our giant models with previous results as the physics included in both sets of models are identical. 

%There are not many studies that have looked at the helium abundance, however, there are many to determine the age. Using white dwarfs, \citet{Bedin2008Reaching67911} found a turnoff age of 8 Gyr, and a white dwarf cooling age of $\sim$6 Gyrs. \citet{Garcia-Berro2010}, using some updated physics, similarly finds the 8 Gyr turnoff age, but determine the cooling age to be $7.0\pm0.3$ Gyr. With global asteroseismic results, \citet{Basu2011} found ages in the range of 6.8 to 8.6 Gyrs, depending on the models used, with a typical error of 0.5 Gyrs. \citet{Carraro2006} fit isochrones, with [Fe/H] = 0.39, from several sources to the cluster and determined the age to likely be between 7.5 and 8.5 Gyrs.

\begin{figure*}
\epsscale{1.1}
\plotone{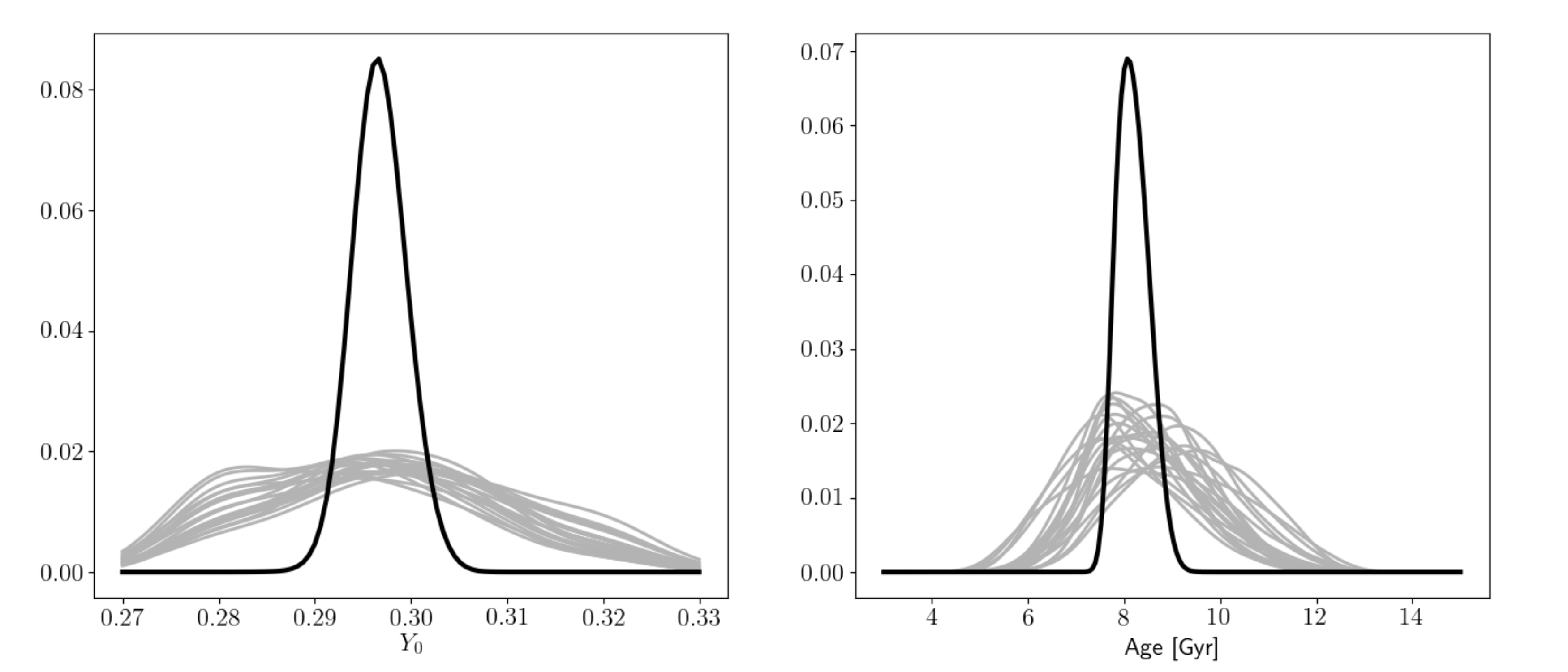}
\caption{One dimensional probability distributions for all targets (individual gray lines) and the combined joint probability distribution (black line) for helium abundance (left) and age (right)   \label{fig:1d_all}}
\clearpage
\end{figure*}

\subsection{Red giants} 
\begin{figure}
\epsscale{1.25}
\plotone{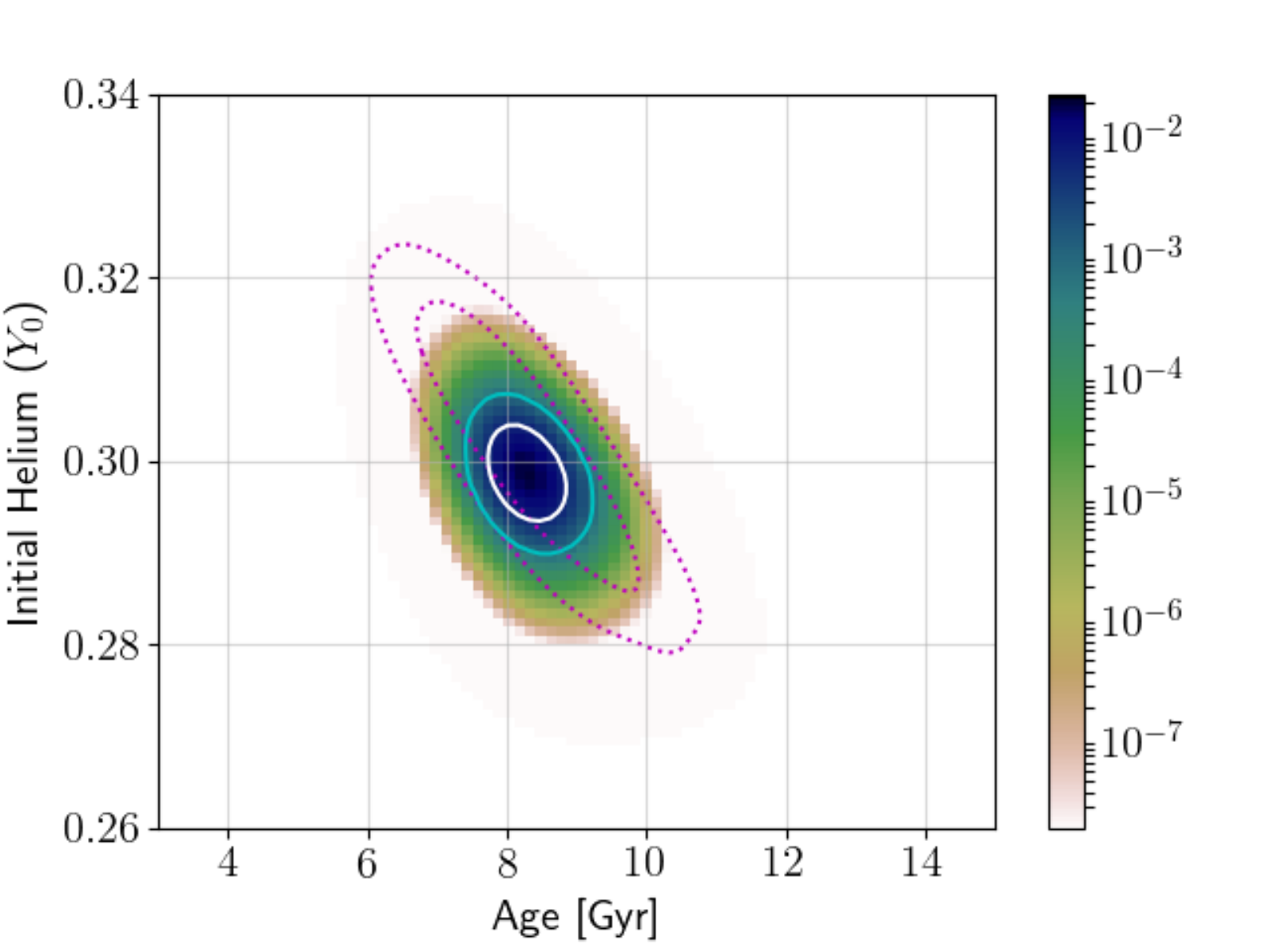}
\caption{Joint probability distribution of age and initial helium abundance for the for the red giants of NGC 6791. The 1-$\sigma$ and 2-$\sigma$ error contours are drawn in white and cyan, respectively. The error contours from the binary analysis (Fig.~\ref{fig:binary_2d}) are shown in magenta dotted lines.\label{fig:rgb_2d}}
\clearpage
\end{figure}

In our analysis of the red giants, we first calculated the probability density distribution of initial helium abundance and age for each red giant in the sample independently. \citet{Valle2018age} noted that for a synthetic sample of cluster stars, based on the properties of NGC 6791, the estimated 1-$\sigma$ error in the age from their grid-based analysis was $\sim$1.7 Gyr, however this estimate does not to consider the sample as a complete, coeval set. As these are all members of a single star cluster, with a single age and composition, we consider the joint distribution of all stars in the sample. Fig.~\ref{fig:1d_all} shows these 1-d distributions of helium abundance and age, as well as the combined joint probability for the entire set of stars.  We are particularly interested in the 2-d joint distribution of age and helium abundance, which can be seen in Fig.~\ref{fig:rgb_2d}. The 1-$\sigma$ and 2-$\sigma$ error contours drawn in white and cyan for this distribution.  For the helium abundance, we find $Y_0 = 0.297\pm0.003$, which is compatible with the 0.30 that B2012 %\citet{Brogaard2012a} 
reported.  We find an age of $8.2\pm0.3$ Gyrs, which is, again, consistent with both the results from the analysis of the eclipsing binaries above, and the results of B2012. By using 27 individually measured stars we are able to reduce the uncertainties on the analysis significantly as compared to using only the four binary stars. This can be clearly seen in Fig.~\ref{fig:rgb_2d}, where we plot the error ellipses from the previous section on top of our results.

%\begin{figure*}
%\epsscale{1.0}
%\plotone{echelles.png}
%\epsscale{1.0}
%\plotone{echelles2.png}
%\caption{Echelle diagrams showing the observed modes (open symbols) and the model frequency (closed symbols) for $l=0$ (circles) and $l=2$ (triangles), for several models near the optimal result for KIC 2437781 (top) and KIC 2436688 (bottom). \label{fig:ech}}
%\end{figure*}

\begin{figure}
\epsscale{1.25}
\plotone{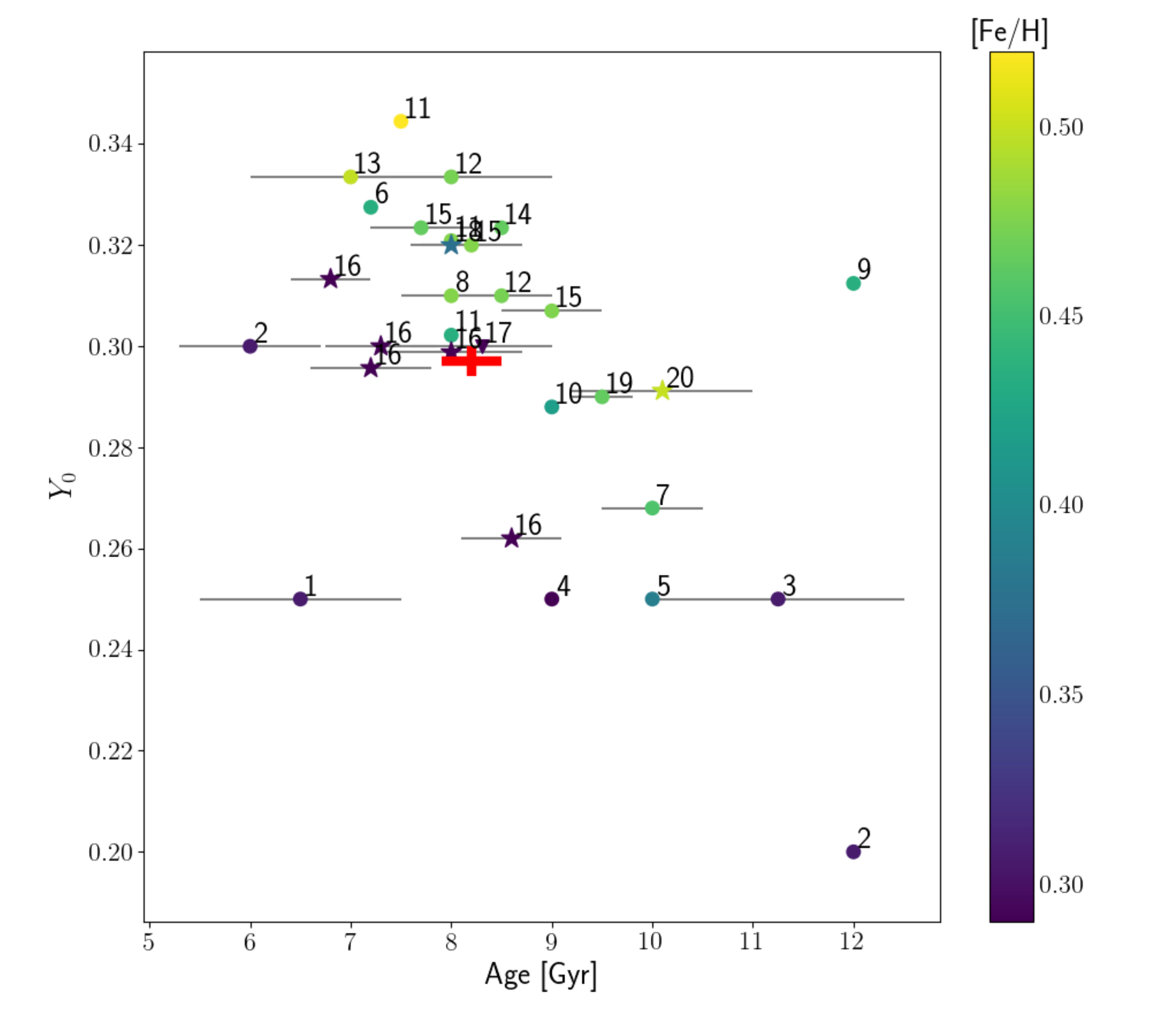}
\caption{Comparison of literature values for age and helium abundance used to derive given age. Our results are shown as the red cross where the width and height are indicative of the size of the error. Each published value is colored by the metallicity. Different symbols reflect the method used to obtain the result: isochrones (circles), binary stars (triangles), and asteroseismology (stars). Numbers correspond to the following references: $^1$\citet{Harris1981}, $^2$\citet{Anthony-Twarog1985}, $^3$\citet{Kaluzny1990}, $^4$\citet{Garnavich1994}, $^5$\citet{Montgomery1994}, $^6$\citet{Kaluzny1995}, $^7$\citet{Tripicco1995}, $^8$\citet{Chaboyer1999}, $^9$\citet{Stetson2003}, $^{10}$\citet{King2005}, $^{11}$\citet{Carney2005}, $^{12}$\citet{Carraro2006}, $^{13}$\citet{Anthony-Twarog2007}, $^{14}$\citet{Kalirai2007}, $^{15}$\citet{Grundahl2008}, $^{16}$\citet{Basu2011}, $^{17}$\citet{Brogaard2012}, $^{18}$\citet{Wu2014Hayashi}, $^{19}$\citet{An2015},  $^{20}$\citet{Kallinger2018}. \label{fig:age_comp}}
\clearpage
\end{figure}

There are not many studies that have looked at the helium abundance specifically, however, there are many to determine the age. Table~2 of \citet{Wu2014DM} summarizes the literature on NGC 6791 very well. To put our values in the context of previous literature we have plotted our values of helium abundance and age against many of the published literature values in Fig.~\ref{fig:age_comp}. Most of the values come from isochrone fitting to various photometric observations, where the isochrones assumed either a given $\Delta Y/ \Delta Z$ relation, or a fixed $Y$ in a few cases. The earlier studies \citep{Harris1981,Anthony-Twarog1985,Kaluzny1990} had metallicities near the solar value and were largely dominated by either Yale isochrones \citep{Ciardullo1977} or those of \citet{Vandenberg1983,Vandenberg1985}, which found drastically different ages between 6 and 12 Gyrs.

More recent work by \citet{Carney2005,Carraro2006,Kalirai2007} and \citep{Anthony-Twarog2007}, which had metallicities closer to [Fe/H] $\sim$ +0.4, all found relatively consistent ages within the range of 7--9 Gyrs, no matter the choice of isochrone. All of these studies had helium abundances that were fixed by their chosen $\Delta Y/ \Delta Z$ relation, and were near the high end ($Y_0>0.30$) of helium abundances found in the literature.

%Using white dwarfs, \citet{Bedin2008} found a turnoff age of 8 Gyr, and a white dwarf cooling age of $\sim$6 Gyrs following BaSTI. \citet{Garcia-Berro2010}, using some updated physics, similarly finds the 8 Gyr turnoff age, but determine the cooling age to be $7.0\pm0.3$ Gyr. 
With global asteroseismic results, \citet{Basu2011} found ages in the range of 6.8 to 8.6 Gyrs from several different sets of isochrones, with a typical error of 0.5 Gyrs. Two of their sets of models were computed for the same set of isochrones \citep[YREC,][]{Demarque2008} with a higher helium abundance of $Y_0=0.30$, and a  regular value of $Y_0=0.26$. The difference in helium abundance had the effect of reducing the age by 1.3 Gyrs in the higher helium models. However, our age result agrees more strongly with the lower helium abundance model. Using the $Y^2$ isochrones \citep{Demarque2004} instead, \citet{Basu2011} find an age of 8.0$\pm$0.7 Gyrs for a helium abundance of $Y_0\sim0.30$, which is in excellent agreement with our work. In contrast to the asteroseismic-based results of \citet{Basu2011}, recent work by \citet{Kallinger2018} determined an age of the cluster using similar grid-based asteroseismic methods that was significantly larger ($10.1\pm0.9$ Gyr) using their non-linear corrections to the scaling relations. An age of 10.1 Gyrs is inconsistent with our joint analysis of all the cluster stars. 

Our analysis also enables us to derive a mean mass of red giant stars of $\overline M_{RGB} = 1.15 \pm 0.01$.  \citet{Basu2011} found a mean mass of red giants of 1.20$\pm$0.01\Msol\ and \citet{Miglio2012} reported $\overline M_{RGB} = 1.23\pm 0.02$. \citet{Miglio2012} applied a small correction to \dnu\ in their analysis before computing masses, however both of these values are significantly higher than our value. This can be partially attributed to our analysis covering a smaller portion of the red giant branch; our analysis covers $\log g\sim2.4$--3.0, while both \citet{Basu2011} and \citet{Miglio2012} use stars further up the giant branch with a lower $\log g$ limit of $\sim$2.1. However, part of the difference can also come from the use of the asteroseismic scaling relations, which have been noted to need some form of a correction \citep{White2011,Gaulme2016} to both \dnu\ and \numax\ \citep{Viani2017}. \citet{Kallinger2018} derived non-linear corrections to the scaling laws, and estimated the mean mass of ascending red giant branch stars to be $1.10\pm0.03$\Msol\ for NGC 6791. This is also in slight disagreement with our value, likely because of the choice of scaling correction applied, but closer than the uncorrected asteroseismic values.  The deviations can partially be attributed to the assumption that \dnu\ follows only a first-order asymptotic expression, which leads to the need for corrections. A summary of several different corrections to the scaling laws, and their applicability in the context of red giants in eclipsing binaries, can be found in \citet{Gaulme2016} and \citet{Brogaard2018rgeb}. Most of the corrections are empirical in nature, but recent work by \cite{Ong2019} show physical motivation for deriving a better estimator of \dnu\ from stellar models.

In B2012, they determined a mass of the lower RGB stars to be $1.15\pm.02$ from their analysis of the main sequence binary stars and chosen isochrones. This value is consistent with our results. A blue straggler system in NGC 6791 was modeled by \cite{Brogaard2018}, who determined the initial masses of the two components. The component who donated mass to the companion during its red giant phase began its life with a mass of 1.15 \Msol, which agrees well with the mean mass of red giants that we studied.

We have assumed here that the NGC 6791 is a single-aged stellar population--a valid assumption for open clusters given the evidence so far \citep{Bragaglia2014,Cunha2015,Villanova2018}.  Globular clusters, on the other hand, have been shown to have multiple stellar populations \citep[][and references therein]{Bastian2018}. The ability to detect and distinguish the multiple populations within globular clusters through asteroseismology could provide some possible insight into the formation and evolution of such clusters, however, the data to do such an analysis at the level of precision required is currently lacking.  The K2 mission observed more that 20 clusters, both globular and open \citep{Dotson2018}, however the length of observations is much shorter ($\sim$80 days) than what we have for NGC 6791, and thus the frequency resolution of modes will be worse. There is an opportunity here for TESS (near the continuous viewing zone) or PLATO observations to provide the data needed for such a project.

\section{Conclusions}
\label{sec:discussion}

It is important to study the helium abundance of stars because it is a significant driver of stellar evolution. Although it is not possible to directly measure the helium abundance through spectroscopy in cool stars, asteroseismology provides a valuable method for studying the helium abundance through the modeling of stellar oscillations. 

In this paper we presented our work on the asteroseismic modeling of red giants in NGC 6791.  We looked at the initial helium abundance and age through stellar models and theoretical oscillation frequencies of those models. By using the information contained in the frequencies of 27 red giant stars we were able to determine the age of NGC 6791 to a higher precision than just using global oscillation properties. In addition, the joint analysis of all stars allows us to present the most precise measures of age and helium abundance for the cluster thus far.

We first modeled two binary star systems in the cluster and find $Y_0 = 0.299\pm0.011$ and an age of $8.5\pm1.1$ Gyr.  We then fitted observed mode frequencies and matched them to theoretical frequencies from a large grid of stellar evolution models. From this we find an initial helium abundance $Y_0 = 0.297\pm0.003$ and an age of $8.2\pm0.3$ Gyr. The ages we find are consistent with several of the several of the recent literature values that used isochrones with helium abundances around 0.30. Finally, we determined the mean mass of first ascent red giant branch stars to be $1.15\pm0.008$\Msol.  

Future observations from missions such as TESS, PLATO, and \textit{Gaia} \citep{GAIA2016,GAIA2018} will provide further data, potentially from other clusters, for which we can do these types of studies. Information from \textit{Gaia}, such as luminosity, is quite complementary to the lightcurves commonly used in asteroseismic analyses. There is much potential for deeper study, in particular, of the helium abundance, beyond what has been done here. An in-depth look at all the mode frequencies can reveal information about acoustic glitches and provide constraints on the depth of the helium ionization zone for evolved stars \citep[][and references therein]{Broomhall2014}. Additionally, the methods of mean density inversions have been extended to more evolved stars, which allows for the further exploitation of information contained in the oscillations \citep{Buldgen2019}. 
%This work provides a measure of helium abundance of 27 red giants within the open cluster NGC 6791, and is an important contribution to our understanding of  the history of NGC 6791, whose old age and rich metallicity provide strong constraints on the evolution of chemicals in the galaxy.

\acknowledgments
We thank the referee for the comments that have improved this manuscript. This work was supported in part by  NSF grant AST-1514676 and NASA grant NNX16AI09G. E.C. is funded by the European Union’s Horizon 2020 research and innovation program under the Marie Sklodowska-Curie grant agreement No. 664931.

\textsl{Software:} MESA, GYRE, \dia

\textsl{Facilities:} \kepler

\bibliography{NGC6791_ads} 
\end{document}